# Ionized Nitrogen Mono-hydride Bands are Identified in the Pre-solar and Carbonado Diamond Spectra

J. GARAI[1,2*]

[1]Department of Mechanical and Materials Engineering, Florida International University, Miami, FL 33199
[2] Department of Mechanical and Structural Engineering, *Ybl Miklós Faculty, Szent István University, Budapest, Hungary*

E-mail: jozsef.garai@fiu.edu

**Abstract**-None of the well established Nitrogen related IR absorption bands, common in synthetic and terrestrial diamonds, have been identified in the pre-solar diamond spectra. In the carbonado diamond spectra only the single nitrogen impurity (C centre) is identified and the assignments of the rest of the nitrogen-related bands are still debated. It is speculated that the unidentified bands in the Nitrogen absorption region are not induced by Nitrogen but rather by Nitrogen-hydrides because in the interstellar environment Nitrogen reacts with Hydrogen and forms $NH^+$; $NH$; $NH_2$; $NH_3$. Among these Hydrides the electronic configuration of $NH^+$ is the closest to Carbon. Thus this ionized Nitrogen-mono-hydride is the best candidate to substitute Carbon in the diamond structure. The bands of the substitutional $NH^+$ defect are deduced by red shifting the irradiation induced $N^+$ bands due to the mass of the additional Hydrogen. The six bands of the $NH^+$ defects are identified in both the pre-solar and the carbonado diamond spectra. The new assignments identify all of the nitrogen-related bands in the spectra, indicating that pre-solar and carbonado diamonds contain only single nitrogen impurities.

## INTRODUCTION

Nitrogen is the most important impurity in diamonds. The content and the type of the nitrogen defects are even used to classify the diamonds (Davis, 1977). If the nitrogen content is significant, typically measured in ppm but can be as high as 1%, then the diamonds are Type I while diamonds with undetectable or very small amounts of nitrogen are Type II. When a diamond forms, nitrogen atoms can replace the carbon atoms, forming an isolated single impurity. Thus newly formed diamonds, like synthetic diamonds, almost exclusively contain single nitrogen impurities. Energetically, a higher level of aggregation of nitrogen impurities is favored, single nitrogen impurities tend to unify and form higher levels of aggregation. The bonds between the carbon atoms in diamonds are very strong resulting in very slow diffusion, even on the time scale of a billion years. Elevated pressures and temperatures increase diffusion rates resulting in higher levels of aggregation, clustering two, four, or more nitrogen atoms. The aggregation level of nitrogen can be used to estimate the mantle residence time for terrestrial diamonds. The lack of higher level nitrogen aggregation indicates that diamonds were not affected by high pressures and temperatures and long annealing times after formation. The nano-size of the pre-solar diamonds, which typically contain in the order of $10^3$ C atoms and 10 N atoms (Jones & d'Hendecourt, 2000; Jones, 2001), might put constraints on the Nitrogen aggregation.

About 98% of natural terrestrial diamonds contain defects with 2 and 4 aggregated nitrogen atoms. Based on the aggregation level of the nitrogen, Type I diamonds are subdivided into two subgroups. Type Ia diamonds (98%) contain clustered nitrogen atoms, and Type Ib (0.1%) contain scattered or single nitrogen atoms. Type II diamonds are also subdivided into Type IIa (1-2%), which is almost pure carbon, and to Type IIb (0.1%) containing boron atoms. The percentages refer to the natural occurrence of different diamond types. The aggregation level of nitrogen can be identified by infrared absorption. The single (C center), pair (A center), four (B center) nitrogen and higher aggregations, like platelets each induce well-defined characteristic infrared absorption bands. These nitrogen-related bands are present in terrestrial and synthetic diamonds but lacking in the pre-solar diamond IR spectra (Lewis et al. 1989; Colangeli et al. 1994; Koike et al. 1995; Mutschke et al. 1995; Hill et al. 1997; Andersen et al. 1998; Braatz et al. 2000).

The spectra of carbonado diamond contain the bands relating to substitutional N defects at 1130 cm$^{-1}$ and 1344 cm$^{-1}$; however, the spectra is very different from type Ib diamonds (Fig. 1) and contains many additional unidentified bands in the Nitrogen region. No other bands relating to higher nitrogen aggregation are present in the spectra. The rest of the characteristics bands identified in the carbonado diamond spectra (Garai et al. 2006; Kagi and Fukura 2008) are similar to the pre-solar bands; therefore, these diamonds are also included in this investigation. It should be noted that the origin of carbonado remains enigmatic. The almost identical IR spectra are not an ultimate proof for same origin and formation. The assignments of the nitrogen-related bands in the pre-solar and carbonado diamonds spectra are still debated and the exact configuration of N is not known.

# NITROGEN IN INTERSTELLAR SPACE

The nitrogen in the Hydrogen rich enviroment of interstellar space can react with Hydrogen forming nitrogen hydrides. The nitrogen molecule $N_2$ can disassociate through the reaction forming $N^+$ (Adams & Smith, 1984)

$He^+ + N_2 \rightarrow (N^+_2)^* + He$

$(N^+_2)^* \rightarrow N^+_2 + h\nu$

$\rightarrow N^+ + N$

The kinetic energy of this reaction is 0.28 eV, which is expected to equally divided between $N^+$ and N. This energy is sufficient to overcome on the endoergicity of reaction

$N^+ + H_2 \rightarrow NH^+ + H.$

Atomic nitrogen can also be ionized by cosmic rays in interstellar space

$N + h\nu \rightarrow N^+ + e^-,$

or

$N + p \rightarrow N^+ + e^- + p$

and form $NH^+$ ion (Mitchell et al. 1978). $NH^+$ can also be formed through charge transfer reaction from NH as:

$H^+ + NH \rightarrow NH^+ + H,$

or through the exothermic reaction:

$H_2^+ + N \rightarrow NH^+ + H,$

or can be synhetized from interstellar ammonia as:

$He^+ + NH_3 \rightarrow NH^+ + H_2 + He,$

or through photolytic reaction (Huebner and Giguere, 1980) as

$NH_3 \rightarrow NH^+ + H_2 + e^-,$

or synthetized from HCO as:

$N^+ + HCO \rightarrow NH^+ + CO,$

or through the ionizing reaction of

$NH + h\nu \rightarrow NH^+ + e^-.$

In Hydrogen rich environment $NH^+_2$ and $NH^+_3$ can be formed as:

$NH^+ + H_2 \rightarrow NH^+_2 + H$ and

$NH^+_2 + H_2 \rightarrow NH^+_3 + H$

In reverse NH and $NH_2$ can also been formed through the photochemical decomposition of $NH_3$ (Kerns & Ducan, , 1972).

The exact mechanisms for formation of nitrogen hydrides in space enviroment are still debated; however, it is widly agreed and supported by observations that significant part of the nitrogen in space is in the form of nitrogen hydrides ($NH^+$; $NH$; $NH_2$; $NH_3$ ). Ammonia $NH_3$, is widely observed in dark clouds and star-forming regions (Cheung et al. 1968). and the presence of nitrogen hidrites ($NH$ and $NH_2$ ) are well known in comets (e.g. Swings et al. 1941; Meier et al. 1998; Feldman et al. 1993). Both hydites, $NH$ and $NH_2$, have also been observed in stellar photospheres (e.g. Schmitt 1969; Farmer & Norton 1989). Interstellar $NH$ was first detected in the diffuse clouds towards the star ξPer and HD 27778 by Meyer & Roth (1991). The presence of $NH$ has also been detected in other regions (Goicoechea et al., 2000; Crawford & Williams, 1997; Hily-Blant et al., 2010; Persson et al., 2010). Significant concentration of $NH^+$ ion in interstellar molecular cloud and in comets has been predicted by Almeida and Singh (1982) and detected by Persson et al. (2010). It can be concluded that in interstellar environment Nitrogen reacts with Hydrogen and present in the form of Nirogen-hydrides $NH^+$; $NH$; $NH_2$; $NH_3$.

**DISCUSSION**

In order to extract the pre-solar diamonds from meteorites, the samples go through severe chemical treatments, which alter the original spectra (Braatz et al. 2000). Based on these contaminations it has been concluded that in the pre-solar diamond IR spectra only the absorption features relating to nitrogen atoms (Braatz et al. 1999) and to the multiple phonon absorption of diamonds (Hill et al. 2000) can be considered as intrinsic properties of the diamond. In this study the absorption features of Nitrogen impurities are investigated.

The complete data set of previous pre-solar and carbonado diamond IR spectra reported in the literature was collected. The characteristic bands are listed in Table 1.

In the Nitrogen region the band at 1085 cm$^{-1}$ has been proposed results from chemical treatment (Braatz et al. 2000). The spectra of carbonado diamond contain all of the reported bands of pre-solar diamonds with the exception of the 1085 cm$^{-1}$. Since the carbonado samples were not chemically treated (Garai et al., 2006), the lack of the 1085 cm$^{-1}$ band is consistent with the proposed chemical treatment-induced origin of this band.

The band at 1285 cm$^{-1}$, characteristic of A center in terrestrial diamonds, is reported in the Orgueil and the carbonado diamond spectra. This band is greatly diminished by further oxidation of the nano-diamond residues of Orgueil chondrite indicating that the band is an artifact resulting from the chemical treatments (Hill et al. 1997). This band is present in all of the chemically treated carbonado diamond spectra (Kagi & Fukura, 2008); however, it is also detected in one of the non-treated carbonado spectra. Thus the possibility that this band might be induced by A centers can not be excluded.

The characteristic bands of C centre at 1130 cm$^{-1}$ and 1344 cm$^{-1}$ are present in the spectra of carbonado diamond. In the Allende and Orgueil spectra there is a band at 1122-1125 but the 1344 cm$^{-1}$ band is missing. Thus the presence of the substitutional Nitrogen impurities (C centre) is debatable in pre-solar diamonds.

None of nitrogen related bands, common in terrestrial and synthetic diamonds, are uniquely identifiable in any of the pre-solar diamond spectra. Based on the lack of the well established Nitrogen related bands, it is speculated that the unidentified bands in the Nitrogen absorption region might not induced by Nitrogen but rather by Nitrogen-hydrides because in the Hydrogen rich interstellar environment Nitrogen reacts with Hydrogen and forms $NH^+$; $NH$; $NH_2$; $NH_3$.

The bonds between N and C are covalent. Covalent bonds are directional and dictates the spatial arrangements of the atoms. Carbon atoms in the diamond structure have tetrahedron coordination based on the spatial distribution of the sp$^3$ hybridized electrons. In order to fit into the diamond lattice the spatial distribution of the electrons of Nitrogen has to be the same. The electronic structure of Nitrogen is $2s^2\ 2p^3$. The three bonding domains of the p electrons and the one nonbonding domain of the loan electron par electrons forms a tetrahedron skeleton structure which fits into the diamond lattice. Thus, the single nitrogen impurity in the diamond lattice is bonded to three of the neighboring carbon atom. This conclusion is supported by the existence of the NV defects. If Nitrogen would bonded to four carbon atoms then the NV defects in diamond would not be stable and NV defects would not exist permanently. The tetrahedron skeleton structure of the diamond lattice can also be achieved through the ionization of the Nitrogen. If Nitrogen is ionized then the remaining four electrons are hybridized to sp$^3$. Thus $N^+$ fits comfortably into the diamond lattice by forming four bonds. It can be predicted that among the available nitrogen hydrides ($NH^+$; $NH$; $NH_2$; $NH_3$) the coordination of $NH^+$ would preferentially fit into the bulk diamond lattice. Thus, it can be predicted that $NH^+$ could substitute carbon in the diamond lattice and form an $NH^+$ defect when nitrogen hydrides are present at the formation of the diamonds.

The absorption features of the $NH^+$ defects are not known. In an attempt to identify the features $NH^+$ defects the bands of $N^+$ defects were redshifted by increasing the mass of Nitrogen with one atomic units due to the mass of the additional Hydrogen.

The ionization of the substitutional N (C center) and conversion into substitutional $N^+$ ($C^+$ center) has been predicted (Lawson & Kanfa 1993) and later experimentally verified by the irradiation of type Ib diamonds . The irradiation-induced $C^+$ center exhibit a sharp peak at 1332 cm$^{-1}$ and broader and weaker peaks at 1115 cm$^{-1}$, 1046 cm$^{-1}$, and 950 cm$^{-1}$ in their spectra (Lawson et al. 1998). The same bands were observed in the spectra of synthetic diamonds after radiation damage by electrons (Collins et al. 1988) along with two additional bands at 1450 cm$^{-1}$ and 1502 cm$^{-1}$ which have been assigned to C-N. All the bands of the C$^+$ center are red shifted in the carbonado-diamond spectra with the exception of the band at 1332 cm$^{-1}$ (Fig. 2-a). It has been shown experimentally that N$^{15}$ isotope doping does not have an effect on the radiation-induced band at 1332 cm$^{-1}$ (Samoilovitch et al. 1975). Thus no red shift is expected for the 1332 cm$^{-1}$ band from the additional mass of Hydrogen. Using a simple force constant model the rest of the irradiation induced N$^+$ bands are red shifted by adding one atomic unit to the mass of Nitrogen, due to the mass of the additional Hydrogen. The calculated absorption bands of the NH$^+$ defects are listed in Table 2. The employed simple force constant model can be used and gives reasonably good results, when the bonds are uniform and spherically symmetric. The permanently positively charged Nitrogen with tetravalent bonds satisfies these requirements.

The agreement between the red shifted substitutional N$^+$ bands and the bands of the pre-solar and carbonado diamond spectra is very good. Based on this good agreement it is suggested that the ionized Nitrogen-mono-hydride complex bonded into the four neighboring carbon atoms and that the bands observed in the Nitrogen absorption region (Table 1) can be assigned to Substitutional NH$^+$ defects.

The presence of NH$^+$ defects in carbonado diamonds is consistent with Hydrogen isotopic composition investigations, which report about 70 ±30 ppm of Hydrogen, relating to the bulk of the diamond (Demeny et al. 2011).

The presence of NH$^+$ defects in pre-solar and carbonado diamonds is also supported by the observed N-H-related bands. Both pre-solar (e.g., Andersen et al. 1998) and carbonado diamond spectra (Fig. 2-b) contain intense N-H-related stretching vibration bands in the region of 3100-3400 cm$^{-1}$ and an N-H bending band at 1650 cm$^{-1}$. These bands are absent or significantly weaker in the spectra of terrestrial and CVD diamonds (e.g., McNamara et al. 1994).

The two steps formation of NH$^+$ defect, ionizing the substitutional N$^0$ or C center by irradiation and then hydrogenating the ionized nitrogen, might also be possible. However, no N$^+$-related bands can be identified in any of the spectra. The complete lack of this "middle stage" defects in this two step formation process indicates that the NH$^+$ defects were most likely formed by one step process. Thus the ionized Nitrogen Mono-hydrides were substituted into the diamond structure at the time of the formation.

The NH$^+$ related bands at 1420 cm$^{-1}$ and 1465 cm$^{-1}$ overlap with C-H deformation bands; therefore, both NH$^+$ and C-H deformation is assigned to these bands.

Despite the similarities between the pre-solar and carbonado diamonds' IR spectra, there are subtle differences. The substitutional $NH^+$ band at 1332 cm$^{-1}$ is reported only in the carbonado spectra. The nitrogen in carbonado-diamond is dominantly present in the form of $NH^+$ defects but substitutional $N^0$-related bands are also detectable in the spectra indicating that both Nitrogen-Hydrides and Nitrogen were available at the time of the diamond formation. The ratio between $N^0$ and $NH^+$ defects in the diamond could be an indicative of the environment in which the diamonds were formed.

## CONCLUSIONS

It is suggested that the unidentified bands observed in the nitrogen region of pre-solar and carbonado diamond IR spectra arise from substitutional $NH^+$ defects. In the carbonado spectra, the $N^0$ bands are also detectable, which indicates that both Nitrogen and Nitrogen mono-hydride were present at the formation of the diamond.

The bands of the substitutional $NH^+$ defect are deduced by red shifting the irradiation-induced $N^+$ bands due to the mass of the additional Hydrogen. The six bands assigned to the $NH^+$ defects are identified in both the pre-solar and the carbonado diamond spectra. With the new assignments, all of the nitrogen-related bands in the spectra are identified and it is shown that pre-solar and carbonado diamonds contain almost exclusively single nitrogen impurities; therefore, they can be classified as Type Ib.

**ACKNOWLEDGEMENTS:** The authors thank to Andriy Durigin and Subrahmanyam Venkata Garimella for the inspiring discussions and for their thoughtful comments. We also like thank to Mike Sukop for reading and commenting the manuscript. This paper benefited from the careful review of Hugh Hill and Anthony P. Jones.

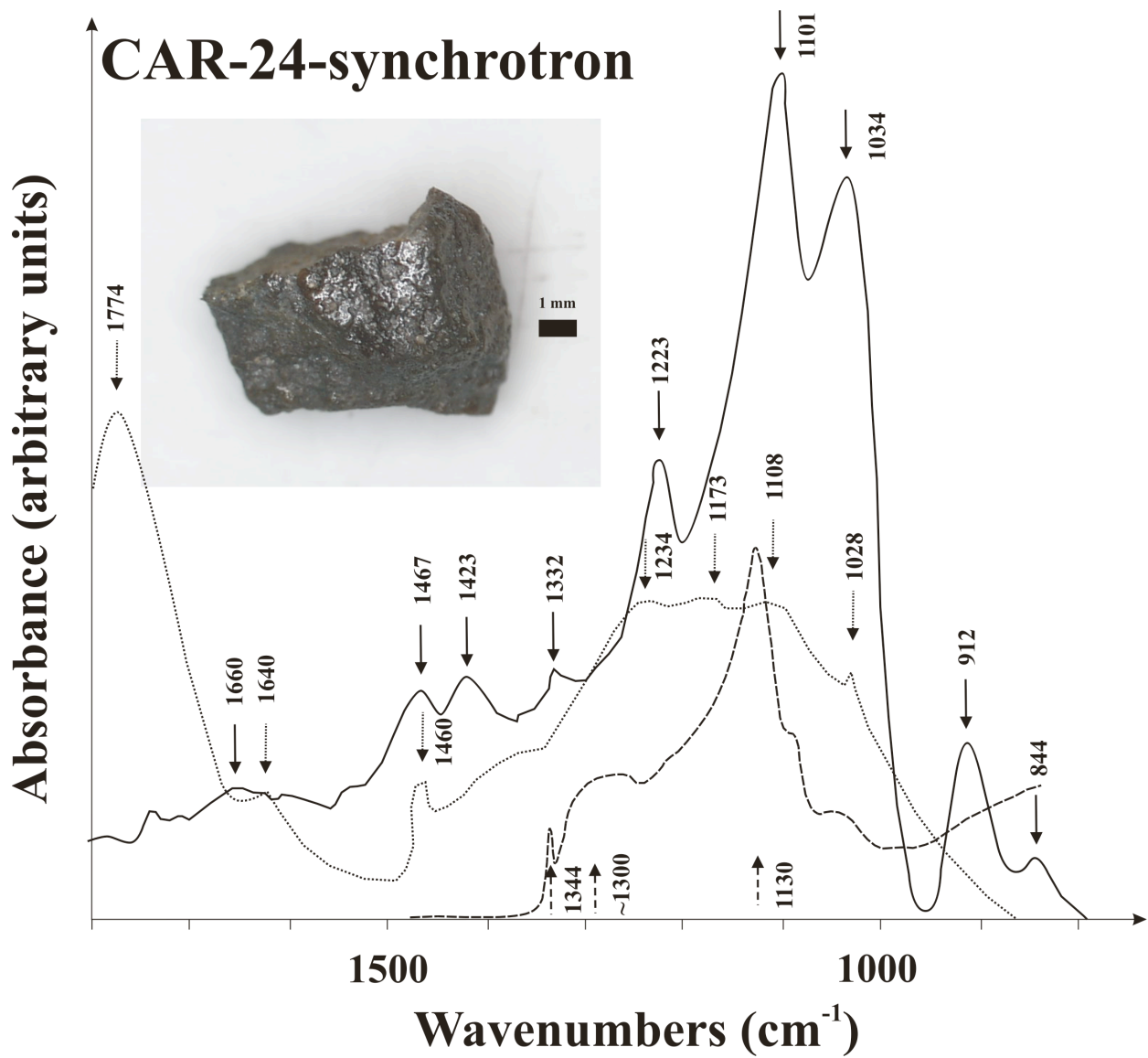

**Fig. 1.** Mid-infrared absorption spectra of carbonado diamond from the Central African Republic (CAR) in the C-N vibrational region (solid line) are plotted. The experimental procedure is described in Garai et al. 2006. Among the characteristic terrestrial diamond absorption bands only the substitutional $N^0$ bands (C Centre) are present in the spectra. The assignments of the rest of the bands are still debated. For comparison the characteristic absorption spectra of Ib diamond containing substitutional N (C Centre) defects (dashed line) (Field 1992) and the spectra of Allende DM1 microdiamond (dotted line) (Lewis et al. 1989) are also shown. The image shows a carbonado from the Bangui Region, Central African Republic. The patina of the surface is a characteristic feature of carbonado diamonds.

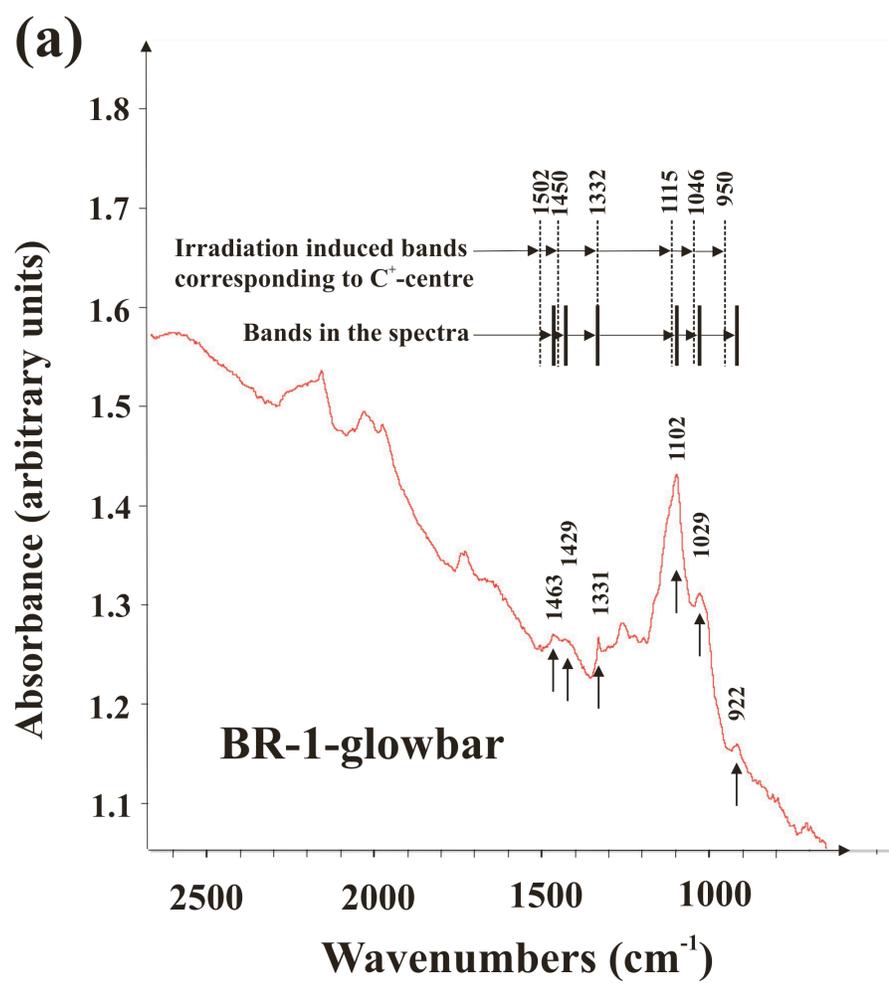

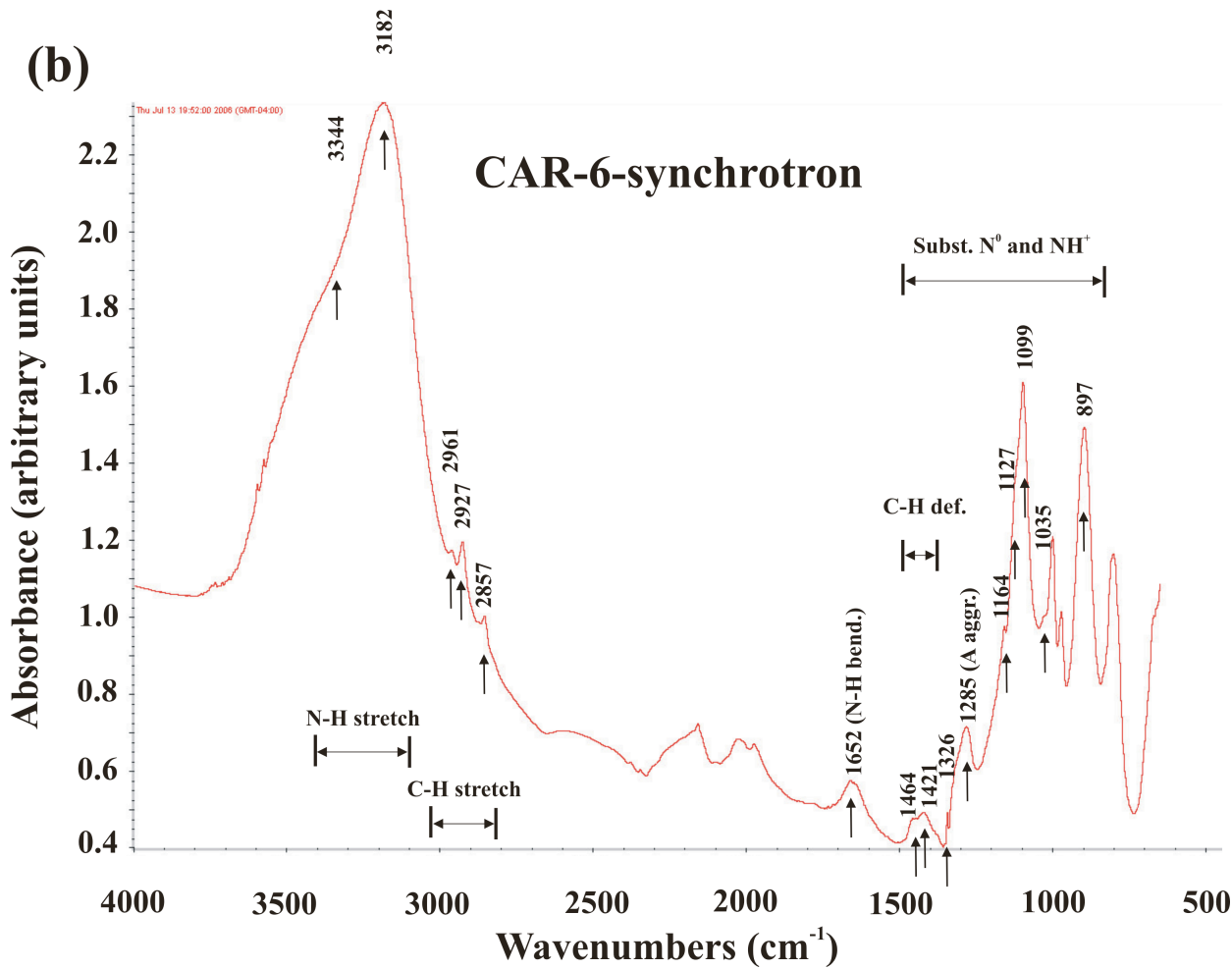

**Fig. 2.** Mid-infrared absorption spectra collected from thin section of Brazilian (BR) and Central African republic (CAR) carbonado diamond are plotted (Garai et al. 2006). The spectra are an intrinsic property of the diamond because no chemical treatment was used in the preparation. The assigned vibrational regions are identified. a) The irradiation-induced $N^+$ related bands are present but red shifted in the carbonado spectra. The observed red shift is consistent with one atomic mass unit increase of the nitrogen, which can be caused by the hydrogenation of the ionized nitrogen. b) The presence of hydrogenated nitrogen is supported by the N-H Stretch and Bending bands at around 3200 cm$^{-1}$ (3.12 μm) and 1652 cm$^{-1}$ (6.05 μm) respectively.

**Table 1** Reported spectral bands for pre-solar and carbonado diamonds in cm$^{-1}$ and their proposed assignments.

| ALLENDE | | | | MURCHISON | | | ORGUEIL | CARBONADO | | | TERRESTRIAL (assignments) | Assignments to presolar and carbonado diamonds by | |
|---|---|---|---|---|---|---|---|---|---|---|---|---|---|
| [1]Lewis et al. 1989 | [2]Koike et al. 1995 | [3]Andersen et al. 1998 | [4]Braatz et al. 2000 | [5]Colangeli et al. 1994 | [6]Mutschke et al. 1995 | [4]Braatz et al. 2000 | [7]Hill et al. 1997 | [8]Garai et al. 2006* | Int. % | [9]Kagi and Fukura 2008 | [10]McNamara and ref. therein 2003 [11]Lawson et al. 1998 | Previous Studies | This Paper |
| 3210 | 3115 | 3236 | | 3164 | | | | (~3200) | | | | [1,2,3,5,6,7] N-H Stretch | N-H Stretch |
| 1640 | 1634 1590 | 1632 | 1612 | 1616 | | 1623 | | 1650 | 26 | 1640 | | [1,5] Aromatic C=C/C=O stretch, [2] C=O/N-H stretching, [3] O-H bend (in $H_2O$) | N-H Bending |
| | | 1462 1456 | | | | | | 1465 | 4 | | | [3] C-H deformation ($CH_3/CH_2$) | Subst. $NH^+$; C-H def.($CH_3/CH_2$) |
| 1403 1361 | 1401 1399 | 1402 1385 | | 1399 | | | | 1420 | 14 | 1438 | | [1,3,5] Interstitial N/ C-H deformation ($CH_3$) | Subst. $NH^+$; C-H def. ($CH_3$) |
| | | | | | | | | (1332) | 3 | 1334 | 1344 (Subst. N)[11] | - | Subst. $NH^+$ |
| | | | | | | | 1289 | | | 1284 | 1282 (A aggregates)[10] | [2,5,6,7] C-O/C-N Stretch | chem. treat.[7]/A cent. |
| 1234 | | | 1210 | | | 1220 | | 1220 | 11 | | 1220 (C-N Stretch)[10] | [3] C-O/C-N/C-C Stretch/ | (C-$N^0$) Stretch |
| 1173 | 1178 | | 1150 | | 1175 | 1155 | | (1163) | 40 | | | Interstitial N | (C-$NH^+$) Stretch |
| | 1122 | 1122 | | | | | 1125 | (1132) | 34 | 1126 | 1130 (Subst. N)[10] | [1] C-O Stretch/ Interstitial N | Subst. $N^0$ |
| 1108 | 1108 | 1109 | | | | | | 1100 | 100 | 1100 | | [4] C-O Stretch | Subst. $NH^+$ |
| 1090 | 1080 | 1090 | 1084 | 1084 | 1090 | 1088 | 1072 | | | | | [4] C-O(H), $CF_2$ | chem. treatment[4] |
| 1028 | | 1054 | | | | | | 1030 | 45 | | | | Subst. $NH^+$ |
| | | | | | | | | 909 | 72 | | | | Subst. $NH^+$ |

*The numbers represent the averages of the bands from the three spectra. The numbers in parentheses relates to bands observed in two of the spectra.

**Table 2** The irradiation induced $N^+$ bands are red shifted by adding one atomic unit to the mass of Nitrogen, due to the mass of the additional Hydrogen. The red shifts were calculated using simple force constant model.

| Assignment | Frequency (cm$^{-1}$) | |
|---|---|---|
| | C-$N^+$ | C-$NH^+$ |
| Irradiation induced Substitutional $N^+$ ($C^+$-centre) | 950 (10.53 μm) | 935 (10.70 μm) |
| | 1046 (9.56 μm) | 1030 (9.71 μm) |
| | 1115 (8.97 μm) | 1098 (9.11 μm) |
| | 1450 (6.90 μm) | 1428 (7.00 μm) |
| | 1502 (6.66 μm) | 1479 (6.76 μm) |